\documentclass[twocolumn]{aastex631}

\begin{document}

\title{Not so fast, not so furious: just magnetic}

\author[0000-0001-8218-8542]{John D. Landstreet}
\affiliation{Armagh Observatory \& Planetarium,
College Hill,
Armagh BT61 9DG, UK}

\affiliation{University of Western Ontario,
1151 Richmond St. N,
London, N6A 3K7, Ontario, Canada}

\author[0000-0003-4936-9418]{Eva Villaver}
\affiliation{Centro de Astrobiolog\'{i}a (CAB), CSIC-INTA, 
Camino Bajo del Castillo s/n, ESAC, 28692, 
Villanueva de la Ca\~{n}ada, Madrid, Spain}

\author[0000-0002-7156-8029]{Stefano Bagnulo}
\affiliation{Armagh Observatory \& Planetarium,
College Hill,
Armagh BT61 9DG, UK}

\begin{abstract}
WD\,0810--353 is a white dwarf within the 20\,pc volume around the Sun. Using {\it Gaia} astrometric distance and proper motions, and a radial velocity derived from {\it Gaia} spectroscopy, it has been predicted that this star will pass within 1\,pc of the Solar System in about 30\,kyr. However, WD\,0810--353 has been also shown to host a magnetic field with strength of the order of 30\,MG. Its spectrum is therefore not like those of normal DA stars of similar effective temperature. We have obtained and analysed new polarised spectra of the star around H$\alpha$. Our analysis suggests that the visible surface of the star shows two regions of different field strength ($\sim 30$ and $\sim 45$\,MG, respectively), and opposite polarity. The spectra do not change over a 4 year time span, meaning that either the stellar rotation period is no shorter than several decades, or that the field is symmetric about the rotation axis. Taking into account magnetic shift and splitting, we obtain an estimate of the radial velocity of the star ($+83 \pm 140$\,km\,s$^{-1}$); we reject both the value an the claimed precision deduced from the {\it Gaia} DR3 spectroscopy ($-373.7 \pm 8.2$\,km\,s$^{-1}$), and we conclude that there will probably be no close encounter between the Solar System and WD\,0810$-$353. We also reject the suggestion that the star is a hypervelocity runaway star, a survivor of a Type Ia Supernova explosion. It is just a stellar remnant in the Solar neighborhood with a very strong and complex magnetic field. 
\end{abstract}

\section{Introduction}\label{Sect_Introduction}
Finding the stars that might have experienced any form of interaction with the Solar System in the past, or those that will do so in the future, is an interesting endeavour which carries important long-term overall implications. A close stellar encounter, or flyby, if it happens with a small impact parameter, has the potential to cause a major disruption in the structure of our planetary system.  But even stellar encounters at larger distances, of the order of 0.5-1\,pc, are expected to be very disruptive, as they can dynamically stir the collection of small bodies that populate the outer Solar System \citep{Wyso2020}. Hypothetically, even those long-range encounters have the potential to affect life on Earth via the temporary enhancement of the cometary influx from the Oort cloud (see e.g. \citealt{Fernandez1987,Dyb2006}). Furthermore, close stellar encounters, along with the interaction with the tidal field of the galaxy \citep{Heisler1986,Portegies2021}, play an important role in the complex chronology of events that lead to the shaping of the Oort cloud  \citep{Oort1950}.

Building the recent history and near future of rubs and scuffs of the Sun with other stars requires an accurate knowledge of current positions and velocities, and the identification of these encounters has proliferated (see e.g. \citealt{Dyb2015,Delafuentemarcos2019,Bailer-Jones2018,Torres2019,Bobylev2020}) as expected with the advent of the {\it Gaia} satellite data \citep{Gaia}, a game changer in this field. 

Recently, the exquisite quality astrometric and photometric data provided by the ESA mission {\it Gaia} has been complemented with the publication in its third data release \citep[DR3][]{Valetal22}. DR3 includes low-resolution (XP) spectral scans based on the BP/RP spectrometers of some 220 million spectra \citep{Deangeli2022,Gaiadr3}. DR3 also reports about 30 million radial velocities obtained with  the Radial Velocity Spectrometer (RVS) by cross-correlating medium-resolution spectra of a short spectral window around the Ca\,{\sc ii} infrared triplet with an appropriate template for each star \citep{Valetal22}. 
The resulting RVs are an extremely valuable source of information for the study of the movement of stars that otherwise do not have any radial velocity (RV) measurement in the literature. This is the case for the star of this paper, the stellar remnant WD\,0810--353, a white dwarf (WD) reported to have a RV of $-373.7\pm 8.2$\,km\,s$^{-1}$  that, among the millions of stars analyzed in the {\it Gaia} DR3, is one of the handful of special stars that might be heading straight towards us in the near future. 

WD\,0810--353 was first identified as a new candidate predicted to experience a close encounter with the Solar System using the {\it Gaia} RV value by \citet{Bobylev2022}. In only $\approx 29$ \,Kyr, the WD is expected to pass at a minimum distance of 0.150$\pm$0.003\,pc from the Sun, the third closest of such identified encounters. This close encounter was discussed again by \cite{Bailer-Jones2022} who flagged it as suspicious based on a possibly incorrect RV determination by {\it Gaia}. The argument not to trust the {\it Gaia} RV measurement was twofold: the expected typical featureless WD IR spectrum (especially in presence of a strong magnetic field), and the fact that the DR3 RV pipeline does not include any WD template. \citet{Delafuentemarcos2022} studied this star in more detail, and considered various interpretations of the feature seen in the {\it Gaia} spectra. Their conclusion was that the star is a likely on a trajectory to approach the Sun in the near future. However, as an alternative, they suggested, on the basis of a weak, possibly blue-shifted H$\alpha$ absorption feature in the XP spectrum, that the WD might instead have a very large radial velocity of about $-4300$\,km\,s$^{-1}$  and thus might be a hypervelocity runaway star. In any case, \citet{Delafuentemarcos2022} called for a new, independent measurement of the radial velocity of the star.
  \begin{figure}
   \centering
   \includegraphics[width=8.7cm,trim={2.0cm 6.3cm 0.7cm 2.5cm},clip]{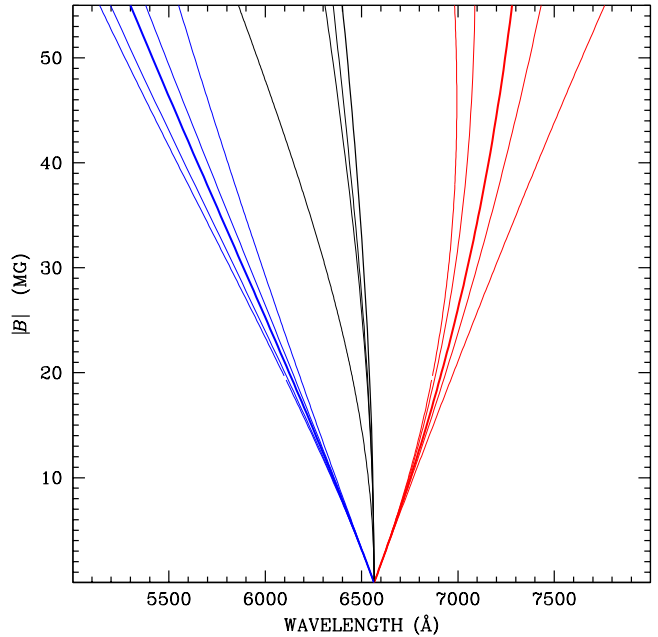}
      \caption{ The spaghetti diagram showing the variations of the wavelengths of components of the Balmer H$\alpha$ as a function of magnetic field strength between 0 and 55\,MG (the thickness of the various lines indicates schematically the relative strengths of the components). Data from \citet{SchWun14}.}
         \label{Spaghetti_diagram}
   \end{figure}

We recall that WD\,0810$-$353 has been clearly identified as a magnetic WD (MWD) on the basis of distinct circular polarisation features seen in the spectrum around 4700--5000\,\AA\ \citep{BagLan20}. Using the blue polarised discovery spectra, it was not possible to model the field morphology  but the star's atmosphere appeared to be hydrogenic, and the field strength was estimated to be of the order of 30\,MG. The presence of such a strong magnetic field has a profound effect on the spectral lines of a star. In a "weak" field of, say, 1\,MG or less, the H Balmer lines split into a simple triplet pattern of a central, undisplaced $\pi$ component flanked by symmetrically displaced $\sigma$ components. In a field of tens of MG or more, however, the splitting of hydrogen lines is far more complex. Each of the $\pi$ and $\sigma$ components of the normal Zeeman effect further splits and shifts, and the wavelengths of the resulting dozen or more components vary strongly with field strength in ways that have been calculated with high precision by the atomic physics community \citep[e.g.][]{SchWun14}. The variations of component wavelengths with field strength are often represented graphically by a {\it spaghetti diagram} like that of Fig.\,\ref{Spaghetti_diagram}. 

In principle, if we can establish accurately the strength of the field affecting one or several spectral line components, and so compute "rest" wavelengths for these features, we can still measure the stellar radial velocity. This requires that we have a sufficiently good knowledge of the magnetic field structure, and are able to identify the field value affecting  discrete, even sharp, features.  As no-one has modelled the magnetic field of WD\,0810--353, we have no reason to assume that these effects were included in the {\it Gaia} RV, and we conclude that {\it Gaia} pipeline RV determination of this object is likely to be incorrect. 

In this paper we present new polarised spectra of WD\,0810$-$353, and derive a qualitative magnetic model of this intrinsically very interesting magnetic WD. Next, we estimate the RV of the star using the guidelines described above, and we re-examine the question of a probable impact parameter of the near future encounter of this star with the Solar System. 

\section{The star}

WD\,0810$-$353 (= UPM\,J0812--3529 = Gaia DR3 5544743925212648320) was discovered to be a nearby WD by \cite{Finch2018}. 
It was subsequently listed as a hydrogen-rich WD in the {\it Gaia} DR2-based  20\,pc sample WD catalogue of \cite{Hollands2018}, who estimated its $T_{\rm eff}$ to be 6217$\pm$9\,K and $\log g = 8.17\pm 0.01$ [dex]. \citet{Jimenez-Esteban2018}  
classified this object as a non-DA, based on VOSA\footnote{http://svo2.cab.inta-csic.es/theory/vosa/}  WD model spectra fits to the SED using synthetic J-PAS photometry derived from the Gaia spectra.  No Sloan Digital Sky Survey (SDSS) spectrum of the target is reported by \citet{Genetal19}, but recently, \cite{Obrien2023} have classified this star as a DC using a Goodman flux spectrum. 
   \begin{figure*}
   \centering
   \includegraphics[width=18.2cm,trim={1.3cm 2.0cm 0.7cm 1.0cm},clip]{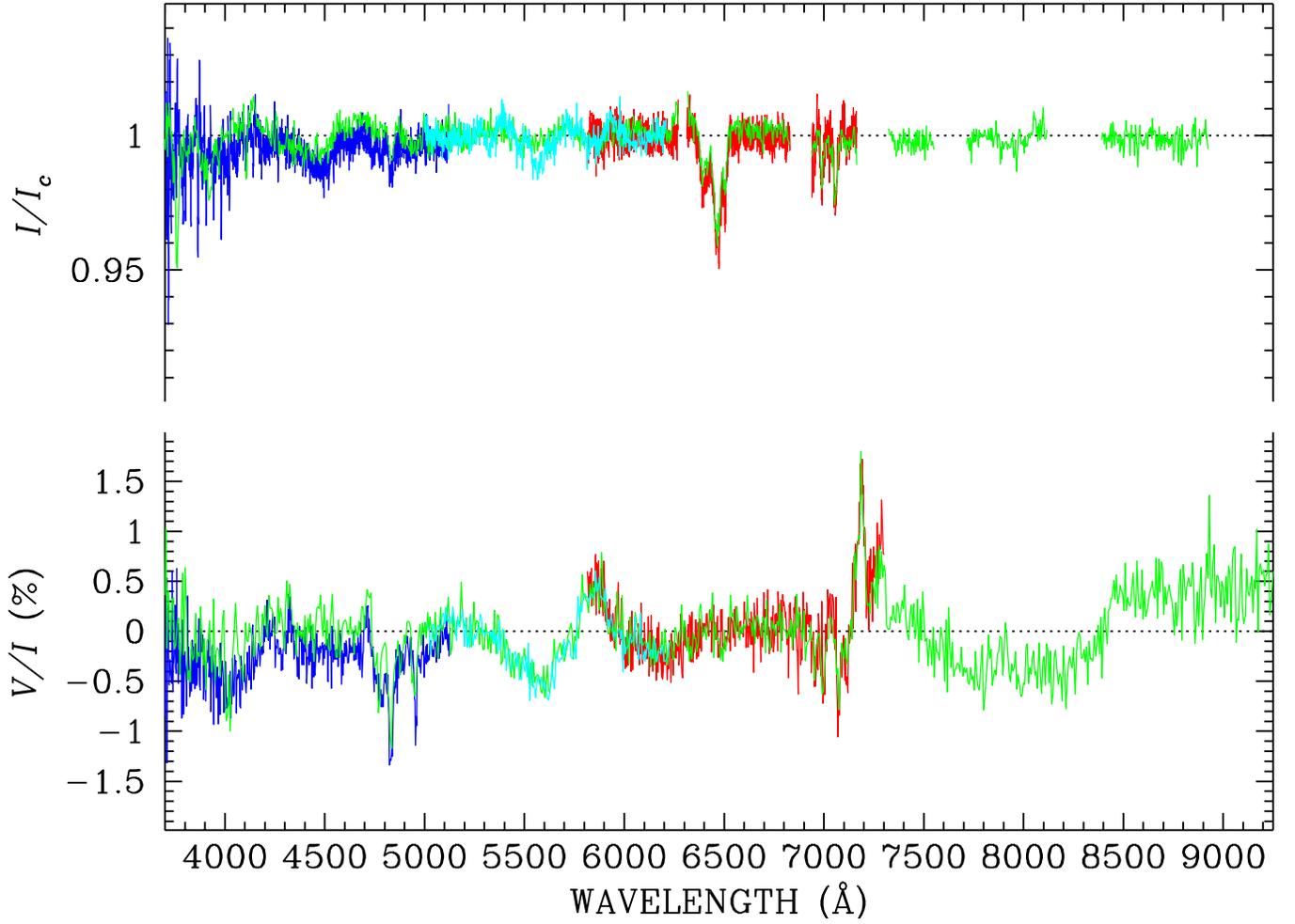}
      \caption{The various $I$ and $V/I$ spectra of WD\,0810--353 obtained with the FORS2 instrument with different settings, as detailed in Table~1. The intensity spectrum was normalised to the pseudo-continuum, ignoring the regions around the major telluric bands. For the sake of clarity, the spectra obtained with grism 600B is plotted only at $\lambda \ga 5000$\,\AA.}
         \label{New_spectra}
   \end{figure*}

Spectropolarimetric data of the star were published by \cite{BagLan20}. These observations covered the range 3700--5100\,\AA\ with a spectral resolving power $R\sim 1200$, and the range $3600-6200$\,\AA\ at $R \sim 600$ (see Table~\ref{Tab_Log}). They revealed a flux spectrum displaying only a few very shallow features, possibly due to flat-field artifacts. However, the circular polarisation spectra reveal firmly detected features of 0.5\% amplitude (but no obvious continuum polarisation). \citet{BagLan20} associated these observed features with the position of two components of the magnetically split H$\beta$ line, and a broad polarisation hump around 5900\,\AA\ with the position of the blue $\sigma$ components of H$\alpha$, concluding that the WD is a DAH star with an atmosphere composed of H. They estimated a field strength of the order of 30\,MG. Figure~6 of \citet{BagLan20}, which displays the variations of the various components of H$\beta$ as a function of field strength \citep{SchWun14}, reveals that with such a large field, virtually none of the magnetically split components of that line are near their zero-field wavelengths.

\section{New observations}
The observations obtained by \citet{BagLan20} were missing most of the region expected to contain the components of the H$\alpha$ line, which is often the easiest line to interpret. Two new circular polarisation spectra were therefore obtained with the FORS2 instrument \citep{Appetal98} of the ESO VLT, one around H$\alpha$ with $R \sim 2100$, and one covering the range $3700-9000$\,\AA\ at $R \sim 440$.  One additional high-resolution spectrum ($R \sim 60,000$) was obtained with the ESPaDOnS instrument of the Canada-France-Hawaii Telescope (CFHT), but its S/N was not high enough to reveal the very weak spectral features, either in flux or in polarisation. All data were obtained and reduced as explained by \citet{BagLan18}, to which we refer the reader also for the definition of the sign of $V/I$. Spectra were aligned with sky lines, and corrected for a gravitational redishift of  $47$\,km\,s$^{-1}$, as well as for heliocentric velocity. The observing log is given in Table~\ref{Tab_Log}. 

The new FORS2 flux and polarisation spectra are shown in Fig\,\ref{New_spectra}, together with those previously obtained by \citet{BagLan20}. All unpolarised flux ($I$) spectra  were divided by spectra of DC WDs obtained with the same setting, then normalised with a polynomial. This normalisation procedure is somewhat arbitrary, and we were not able to assess whether some shallow and broad feature are flatfield artifacts, or the effect of a strong magnetic field on the spectral energy distribution. For example, with our polynomial fitting we removed a broad but shallow flux depression between 5500 and 6500\,\AA, which may be in fact associated with the non-zero circular polarisation observed between 5300 and 6300\,\AA.  The aim of our flux normalisation was to enhance at least the strongest absorption features, particularly from 6375 to 6525\,\AA, and from about 6975 to at least 7100\,\AA\ (in between the strong O$_2$ B band from 6865 to 6935\,\AA\ and the H$_2$O $\alpha$ band starting at about 7160\,\AA). The $V/I$ spectrum shows various structures, which are consistently seen in spectra obtained with different settings (in the common regions). Thanks to the use of the beam-swapping technique \citep[e.g.][]{Bagetal09}, $V/I$ has virtually no normalisation or flat-fielding problems; furthermore, it is largely unaffected by atmospheric absorption bands, except for the decreased S/N (because the Earth atmosphere does not polarise the radiation in transmission). There is little doubt that all polarisation features seen in our spectra are real and intrinsic to the star.

The new low-resolution spectrum taken with the 300V grism allow us to assess the {\it Gaia} RVS radial velocity of the star based on template matching to three lines of the Ca\,{\sc ii} infrared triplet at 8498, 8542 and 8662\,\AA. In our spectrum there are no features whatever anywhere near the positions of these three lines. The flux spectrum is slightly noisy (at the level of $\sim 1$\,\% or less), perhaps due to very weak hydrogen Paschen lines shifted into this region by magnetic splitting. It appears that the RVS template matching procedure probably settled on weak noise features to produce the RV value, and that the reported value of $-373.7 \pm 8.2$\,km\,s$^{-1}$ is definitely spurious.

\begin{table*}
\caption{\label{Tab_Log} Log of the spectropolarimetric observations of WD\,0810-353.}
\begin{center}
\begin{tabular}{lcrclrlrll}
\hline\hline
DATE       & UT    & Exp  &   S/N & Instrument& Grism & Spectral    &$R$ & Reference &Color code      \\
yyyy-dd-mm & hh:mm & (s)  &per \AA&           &       &range (\AA)  &    &           &in Fig.~\ref{New_spectra}\\
\hline
2019-03-24 & 01:14 & 2400 &   470 & FORS2    & 1200B & 3700--5100 & 1400 & \citet{BagLan20} & \\
2020-01-08 & 03:03 & 3680 &   520 & FORS2    & 1200B & 3700--5100 & 1400 & \citet{BagLan20} &blue\\
2020-01-08 & 04:18 & 3600 &   440 & FORS2    &  600B & 3700--6200 &  800 & \citet{BagLan20} &light blue\\
2019-03-21 & 06:32 & 4188 &   146 & ESPaDOnS &       & 3790--9200 &60000 & this work        & \\
2023-01-14 & 04:58 & 2400 &   365 & FORS2    & 1200R & 5700--7100 & 2100 & this work        &red \\
2023-01-14 & 05:31 &  900 &   290 & FORS2    &  300V & 3700--9000 &  440 & this work        & green \\ 
\hline
\end{tabular}
\end{center}
\end{table*}

\section{The magnetic field }

   \begin{figure}
   \centering
   \includegraphics[width=8.5cm,trim={1.5cm 2.5cm 0.7cm 1.0cm},clip]{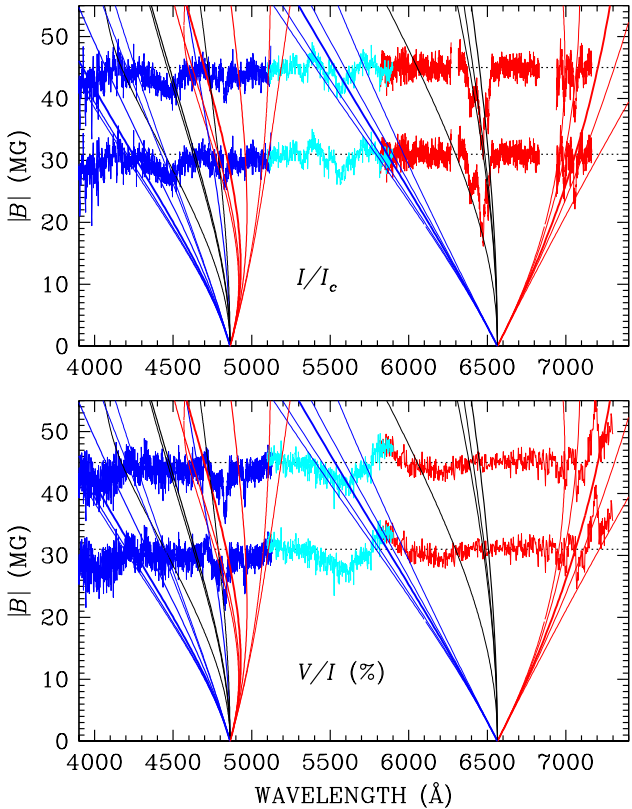}
      \caption{ The wavelengths of the components of H$\beta$ and H$\alpha$ as computed by \citet{SchWun14} compared to the spectra of WD\,0810--353. {\it Top panel:} the normalised $I$ spectrum is overplotted twice, once with the continuum offset to correspond with the 31\,MG level  and once with the 45\,MG level. This shows the coincidence between the computed wavelength of the various line components and the observed absorption features.  {\it Bottom panel:} same as top panel, but for the  $V/I$ spectrum. (Note that the WD polarisation spectrum is not affected by atmospheric absorption bands.)          }
         \label{Absorption_features}
   \end{figure}

The new FORS2 spectra assist us to obtain a clearer view of the stellar surface magnetic field. 

First, by comparing the overlap regions of the new spectra and those obtained in 2019, we see that there is no evidence for variability either in $I$ or in $V/I$ on a time-scale of $\sim 4$ yrs. Therefore, the field is either symmetric about the star's rotation axis, or the stellar rotation period is at least many decades long, a situation found for some other relatively old large-field MWDs such as Grw+70$^\circ$\,8247 \citep[e.g.][]{BagLan19a}. 

Next, consider the weak, complex flux absorption feature between about 6375 and 6525\,\AA, which has no corresponding polarisation feature. The previous estimate of the field strength of the order of 30\,MG by \citet{BagLan20} suggests that this feature is produced by three strong $\pi$ components of H$\alpha$.
The breadth of the observed flux feature hints at a fairly large spread in field strength over the visible hemisphere, running from roughly 30 to 45\,MG. This interpretation rules out the possibility suggested by \citet{Delafuentemarcos2022} that the star has a radial velocity of the order of $-4300$\,km\,s$^{-1}$, which they obtained by assuming that the broad absorption feature in the low resolution {\it Gaia} XP scan of WD\,0810--353, corresponding to the H$\alpha$ $\pi$ components, is due to extreme Doppler shifting of a weak H$\alpha$ line. 

For a field of the order of 25--50\,MG, Fig.~\ref{Spaghetti_diagram} predicts the presence of the blue $\sigma$ components of H$\alpha$ at $\sim 5700$\,\AA, and of the red $\sigma$  components at $\sim 7150$\,\AA. Assuming a smooth magnetic morphology, like that of a centred dipolar field, these $\sigma$ components should have significant $V/I$ signatures of opposite sign. However, the polarisation features of each group of $\sigma$ components show both senses of polarisation at slightly different wavelengths. More specifically, the $V/I$ spectrum shows a broad S-wave around 5700\,\AA, and a more compact and complex S-wave around 7150\,\AA. {\em The simplest interpretation of these features is that they reveal the existence of two areas on the WD surface of opposite polarity and somewhat different field strength.} The smooth positive polarisation bump between 5730 and 5960\,\AA\ is paired with three negative features between 6930 and 7140\,\AA, suggesting the presence of a region of positive longitudinal field (emerging flux lines) and mean field modulus $\sim 30$\,MG. The broad negative $V/I$ feature around 5600\,\AA\ is paired with the strong sharp positive $V/I$ feature at 7190\,\AA. The greater separation of these two features suggests that they reveal a region with field lines entering the star, and a larger field strength of $\sim 45$\,MG.

In Figs.\,\ref{Absorption_features} we explore this preliminary modelling more quantitatively. In the top panel, we again show the positions of the components of H$\alpha$ and H$\beta$ as a function of field strength, just as in Fig.\,\ref{Spaghetti_diagram}. 
Superposed on this spaghetti diagram are two copies of the flux spectrum, one with the continuum offset at the level of $\sim 31$\,MG, and one with the continuum at the level of $\sim 45$\,MG. It can be seen that the assumption that the WD has one large region characterised by a field of $\sim 31$\,MG, and one with a field of roughly 45\,MG, predicts strong magnetic $\pi$ (non polarised) components over the full width of the observed absorption feature around 6450\,\AA. The weak absorption lines sandwiched between the O$_2$ B band and the H$_2$O $\alpha$ band also coincide with red $\sigma$ components produced in the two main field strength areas. 

Next, the bottom panel compares the Schimeczek-Wunner spaghetti diagram to polarisation spectra. This plot shows two copies of the $V/I$ spectrum, one with its (zero) continuum offset to the level of 31\,MG, and one offset to the level of 45\,MG. 

The $V/I$ spectrum offset to 31\,MG shows that the theoretical components of H$\alpha$ coincide with a positive blue bump around 5800\,\AA, and with four negative features close to 7000\,\AA. The lack of individual features in the blue $\sigma$ components arises because there is a significant spread in field strength in the region of positive field, which broadens the effect of each component, as shown by the rapid change with field strength of each of the blue magnetic $\sigma$ components of H$\alpha$. In the red, the rate of change of wavelength of the $\sigma$ components is much slower, and the limited range of ${|B|}$ in this region allows individual components to produce sharper absorption line-like features. Notice that the red components of H$\beta$ also coincide with line-like features of the same sign around 4800\,\AA. The copy of the $V/I$ spectrum placed at a level of about 45\,MG shows that computed $\sigma$ components correspond to the broad negative depression around 5600\,\AA, and to the two strong line-like features in the region of 7200--7300\,\AA. 

These two good matches thus reveal the presence of a region of field strength around 30 to 31\,MG on the visible surface from which field lines emerge, and, somewhere else on the surface, of another region of field strength around $40$ to $45$\,MG, with inward pointing field lines. 

Without detailed polarised spectral modelling, for which we currently do not have adequate tools, we cannot provide a more detailed description of the WD surface field. A further obstacle to a more accurate modelling is that either the star does not rotate, or the field is symmetric about the rotation axis. Either way, this prevents us from seeing the field morphology from different viewing points.

\section{The stellar radial velocity}

We next consider with what accuracy it is possible to estimate the radial velocity of the WD. All the features that one could use for this purpose have wavelengths that depend on the exact value of local field strength, which has substantial dispersion over the visible stellar disk. In the range between 30 and 45\,MG, the wavelengths of the blue $\sigma$ components of the lines change very rapidly with field strength, leading to very broad, blended features. In comparison, the $\pi$ and the red $\sigma$ line components vary more slowly with the field strength. Therefore we will concentrate on these components. 
Specifically, the $\pi$ absorption line in the flux spectrum appears to have clear edges, which reflect the limited range in field strength in the two magnetic regions identified on the visible surface of WD\,0810--353. Similarly, there are also a few red $\sigma$ component absorption lines and polarisation features that have reasonably well-defined edges. We can simultaneously match the positions of the blue edge of the $\pi$ absorption line and the red limit of one or more $\sigma$ components by varying both the upper limit of the field present in the stronger-field region and the RV shift of the spectrum simultaneously. The same exercise can be carried out for the weaker-field region using the red edge of the $\pi$ absorption line and the blue edge of one or more $\sigma$ features to determine simultaneously the strength in the weaker field region and again the RV shift. This procedure provides two independent measurements of the stellar RV that {\it a posteriori} we found consistent with each other.

We carried out the measurements described above, by superposing copies of the flux and polarisation spectra on a plot of the Schimeczek-Wunner spaghetti diagram of component wavelengths as a function of field strength,  shifting these spectra vertically (to optimise field strength limits) and shifting (and stretching) them horizontally (to optimise radial velocity shift). Because the edges of the features used are noisy, this procedure has been carried out visually. Uncertainties were also estimated visually by varying the positions of the fitted features until the fit is judged unsatisfactory. 

We made our measurements on the FORS 1200R spectrum. Using the red-most edge of the broad $\pi$ component of H$\alpha$ near 6510\,\AA, and the blue edge of the strongest of the red $\sigma$ components near 7550\,\AA, we confirmed that the weakest field strength is approximately $30 \pm 0.5$\,MG, while we found  $RV = 183 \pm 200$\,km\,s$^{-1}$. Using the blue edge of the apparent $\pi$ $I$ absorption component at about  6375\,\AA\ and the red edge of the strongest $\sigma$ $V$ component at 7210\,\AA\ we estimated that the strongest field present is about $45.75 \pm 0.5$\,MG, and the corresponding radial velocity is $-17 \pm 200$\,km\,s$^{-1}$. The average of these two measurements is $ RV = +83 \pm 140$\,km\,s$^{-1}$.

The two measurements support our earlier conclusions about the range of local field strengths on the visible stellar surface, and yield two independent measurements of the RV of the star in satisfactory agreement with one another. The mean RV determined from our measurements is, not surprisingly, inconsistent with the large negative RV of $-373.7\pm 8.2$\,km\,s$^{-1}$ reported on the basis of the {\it Gaia} RVS spectrum of WD\,0810--353, and our large uncertainties are consistent with the questions that have been raised about the reported precision of that measurement. 

\section{Conclusions}\label{Sect_Conclusions}
We have obtained and analysed the polarised spectrum of WD\,0810--353, a strongly magnetic cool WD in the solar neighborough. This spectrum shows complex weak flux absorption and polarisation features throughout most of the optical range. The H$\alpha$ $\pi$ component, which is not polarised, is magnetically shifted to the blue, between 6370 and 6520\,\AA, where it appears as a broad and very shallow feature. The H$\alpha$ $\sigma$ components are shifted by hundreds of \AA\ to the blue and to the red. The higher Balmer lines, significantly weaker than H$\alpha$, are similarly split and shifted, leading to further complex weak features in $I$ and $V$. We have interpreted these weak features as produced by the atmosphere of a strongly magnetic DA WD, with two important regions of overall opposite field line polarity, and typical strengths of about 31 (outward field) and 45 MG (inward field). 

Simultaneously to our magnetic modelling, we have also estimated the star's radial velocity ($+83 \pm 140$\,km\,s$^{-1}$). The large uncertainty in our determination of the radial velocity is due to the fact that the dispersion of the magnetic field strength of WD\,0810--353 leads to large and poorly defined dispersion in the wavelengths of all absorption and polarisation features. Our result is in strong disagreement with that obtained in previous works ($ RV = -373.7\pm 8.2$\,km\,s$^{-1}$), which had led to the erroneously precise conclusion that  WD\,0810--353 will pass within a fraction of a parsec of the Solar System within $\sim 30$\,kyr \citep{Bobylev2022}. Furthermore, we are able to rule out the possibility that the strongly blue-shifted $\pi$ component of H$\alpha$ indicates that this star has a huge radial velocity of the order of $-4300$\,km\,s$^{-1}$ \citep{Delafuentemarcos2022}.

Nevertheless, WD\,810--353 is an intrinsically very interesting star. It is one of the closest strongly magnetic WDs to the Earth. With an age of almost 3\,Gyr, it is entering the phase of its cooling life during which very strong magnetic fields emerge to the surface of middle aged-WDs \citep{BagLan21,BagLan22}. It appears to have a fairly complex distribution of local field strength over the visible surface. It will certainly be worthwhile to carry out further spectropolarimetric monitoring and more detailed modelling of this object. 

\vspace{5mm}
\section{Acknowledgements}
\begin{small}
\noindent
Based on observations obtained with data collected at the Paranal Observatory under program ID 110.25A1.001,  and with the ESPaDOnS instrument on the Canada-France-Hawaii Telescope (CFHT) (operated by the National Research Council (NRC) of Canada, the Institut National des Sciences de l’Univers of the Centre National de la Recherche Scientifique (CNRS) of France, and the University of Hawaii). All raw data and calibrations of FORS2 and ESPaDOnS data are available at the observatory archives: ESO archive at {\tt archive.eso.org} and the Canadian Astronomical Data Centre at {\tt https://www.cadc-ccda.hia-iha.nrc-cnrc.gc.ca/en/}. EV acknowledges support from the DISCOBOLO project funded by the Spanish Ministerio de Ciencia, Innovaci{\'o}n y Universidades under grant PID2021-127289NB-I00.
JDL acknowledges the financial support of the Natural Sciences and Engineering Research Council of Canada (NSERC), funding reference number 6377-2016. This research was supported in part by the National Science Foundation under Grant No. NSF PHY-1748958.
We thank the Kavli Institute for Theoretical Physics (KITP) for hosting the program "White Dwarfs as Probes of the Evolution of Planets, Stars, the Milky Way and the Expanding Universe". 
\end{small}

\bibliography{sbabib}{}
\bibliographystyle{aasjournal}
\end{document}